\documentclass[twocolumn,prl,amsmath,amssymb,showpacs,superscriptaddress]
{revtex4-1}
\usepackage{epsf}      
\usepackage{graphicx}
\usepackage{color}
\usepackage{soul}
\usepackage{gensymb}

\begin{document}

\title {Structural, transport, optical and electronic properties of Sr$_2$CoNbO$_6$ thin films}
\author{Ajay Kumar}
\affiliation{Department of Physics, Indian Institute of Technology Delhi, Hauz Khas, New Delhi-110016, India}
\author{Rishabh Shukla}
\affiliation{Department of Physics, Indian Institute of Technology Delhi, Hauz Khas, New Delhi-110016, India}
\author{Akhilesh Pandey}
\affiliation{Solid State Physics Laboratory, DRDO, Lucknow Road, Timarpur, Delhi-110054, India}
\author{Sandeep Dalal}
\affiliation{Solid State Physics Laboratory, DRDO, Lucknow Road, Timarpur, Delhi-110054, India}
\author{M. Miryala}
\affiliation{Shibaura Institute of Technology, Toyosu campus, Koto-ku Tokyo, Japan}
\author{K. Ueno}
\affiliation{Shibaura Institute of Technology, Toyosu campus, Koto-ku Tokyo, Japan}
\author{M. Murakami}
\affiliation{Shibaura Institute of Technology, Toyosu campus, Koto-ku Tokyo, Japan}
\author{R. S. Dhaka}
\email{rsdhaka@physics.iitd.ac.in}
\affiliation{Department of Physics, Indian Institute of Technology Delhi, Hauz Khas, New Delhi-110016, India}
\date{\today}

\begin{abstract}

We study the effect of substrate induced strain on the structural, transport, optical and electronic properties of Sr$_2$CoNbO$_6$ double perovskite thin films. The reciprocal space mapping, $\phi$-scan and high-resolution $\theta$-2$\theta$ scans of x-ray diffraction patterns suggest the epitaxial nature and high-quality of the films deposited on various single crystal ceramic substrates. A systematic enhancement in the dc electronic conductivity is observed with increase in the compressive strain, while a sharp reduction in case of tensile strain, which are further supported by change in the activation energy and density of states near the Fermi level. The optical band gap extracted from two distinct absorption bands, observed in the visible-near infrared spectroscopy show a non-monotonic behavior in case of compressive strain while significant enhancement with tensile strain. Unlike the bulk Sr$_2$CoNbO$_6$ (Co$^{3+}$ and Nb$^{5+}$), we observe different valence states of Co namely 2+, 3+ and 4+, and tetravalent Nb (4$d^1$) in the x-ray photoemission spectroscopy measurements. Moreover, a reduction in the average oxygen valency with the compressive strain due to enhancement in the covalent character of Co/Nb--O bond is evident. Interestingly, we observe sharp Raman active modes in these thin films, which indicates a significant enhancement in structural ordering as compared to the bulk.
\end{abstract}

\maketitle
\section{\noindent ~Introduction}

Tailoring the physical properties of transition metal oxides through the strain engineering in case of epitaxial thin films has become the versatile technique and show unusual phenomenon at the interface in nanoscale range \cite{Ohtomo_Nature_04, Fuchs_PRB_07, Fuchs_PRB_05}. Unlike in their bulk counterpart, the strain caused by lattice mismatch between thin film and substrate can induce the novel phenomenon at the interface \cite{Ohtomo_Nature_04, Fuchs_PRB_07, Fuchs_PRB_05, Locquet_Nature_98, Haeni_Nature_04, Lee_Nature_10} which play a crucial role in device applications \cite{Mannhart10} such as solid oxide fuel cells \cite{KubicekACSN13}. A close synchronization between the substrate induced strain and oxygen non-stoichiometry has been recently established in these compounds, where controlling such parameters in the bulk form is a major challenge \cite{Herklotz_JPCM_17, Petrie_AFM_16, Aschauer_PRB_13}. In this family, perovskite oxides of  type ABO$_3$ (A: rare earth/alkali earth metals, B: transition metals) have been widely explored due to their stable structure and availability of the wide range of structurally alike single crystalline substrates for the heteroepitaxial growth. The misfit induced strain in these compounds is widely known to modify the BO$_6$ octahedron i.e. change in B--O--B bond angle and B--O bond length, where degree and direction of the octahedral rotation is determined by the magnitude and sign of the biaxial strain \cite{Vailionis_PRB_11}. This, octahedral distortion effectively perturb the energy scales of lattice, spin, charge, and orbital degrees of freedom, that control the effective correlation (U/W) between them and alter most of their electronic and magnetic properties \cite{Dhaka_PRB_15, Rondinelli_AM_11}. For example, electronic bandwidth can be written to the first approximation as W$\propto$ cos$\phi$/$d$$^{3.5}$, where $\phi$= ($\pi$-$\theta$)/2 is the buckling deviation of B--O--B bond angle $\theta$ from 180$\degree$ and $d$ is the B--O bond length \cite{Medarde_PRB_95}. The substrate induced compressive strain is known to reduce the electronic band gap due to enhancement in the strength of hybridization between transition metal $d$ and O 2$p$ orbitals while opposite trend is expected in case of tensile strain due to the suppression of this $p-d$ hybridization \cite{Huang_APL_19}.

In recent years, lattice misfit induced strain and nanostructures have been extensively used as an effective tool in the Co-based perovskite oxides to tune the valence and spin states of Co and hence their electronic and magnetic properties \cite{RaviJALCOM18, Choi_NL_12, Mehta_PRB_15, Callori_PRBR_15, Zhang_PRBR_19, Lee_PRL_11}. This is because the strong competition between the crystal field strength ($\Delta_{cf}$) and Hund's exchange energy ($\Delta_{ex}$) results in the considerably small energy difference between low and excited spin states of Co and perturbation produced by the biaxial strain can effectively alter their relative population as compare to the bulk \cite{Choi_NL_12, Mehta_PRB_15, Callori_PRBR_15, Zhang_PRBR_19, Lee_PRL_11, ShuklaPRB18, ShuklaJPCC19}. For example, large epitaxial tensile strain in case of SrCoO$_3$ stabilizes the intermediate spin (IS) state of Co$^{4+}$ (d$^5$, t$_{2g}^4$e$_g^1$, S=3/2) into low spin (LS) state (t$_{2g}^5$e$_g^0$, S=1/2), which induce the antiferromagnetism and metal-insulator transition in the ferromagnetic metallic ground state \cite{Callori_PRBR_15, Lee_PRL_11, Long_JPCM_11}. Interestingly, 50\% substitution of B (Co) atoms by another transition metal atoms (B$^{\prime}$) in these compounds gives an extra degree of freedom to tune the rock salt like ordering of their (B/B$^{\prime}$)O$_6$ octahedra, which can be controlled by the ionic radii and valence mismatch between them \cite{Anderson_SSC_93}. Thus, Co-based ordered perovskites (double perovskites with general formula A$_2$BB$^\prime$O$_6$) give rise to the exotic magnetic, electronic and transport properties due to combined effects of valence and spin state transition of Co and flexibility in tuning the degree of B-site cationic ordering, where electronic band structure can be manipulated through the mechanical, chemical (doping) or misfit induced biaxial strain \cite{Narayanan_PRB_10, Bos_PRB2_04, Bos_CM_04, Xing_JAC_19, Esser_PRB_18, Kleibeuker_AsiaM_17}. For example, electronic band gap has shown to be systematically suppressed with Sr concentration in La$_{2-x}$Sr$_x$CoMnO$_6$ due to enhancement in the concentration of Co$^{3+}$ and hence evolution of new Co$^{3+}$--O$^{2-}$--Mn$^{4+}$ conduction channel \cite{Xing_JAC_19}. Hole doping in Ir t$_{2g}$ band of La$_{2-x}$Sr$_x$CoIrO$_6$ ($0\leqslant x\leqslant$2) for $x\leqslant$1.5 and in Co 3d band for $x\geqslant$1.5 also suppress electronic band gap \cite{ Narayanan_PRB_10}. On the other hand, a similar impact on the electronic band gap has been achieved through the misfit induced biaxial strain in the epitaxial thin films of Sr$_2$CoIrO$_6$ double perovskite \cite{Esser_PRB_18}.
 
In this context, Sr$_2$CoNbO$_6$ is of particular interest due to its interesting structural, magnetic, transport and electronic properties and hence their wide range of applications \cite{Kobayashi_JPSJ_12, Yoshii_JAC_2000, Xu_JCSJ_16, Wang_AIP_13, Bashir_SSS_11, Azcondo_Dalton_15, Ramirez_IJMPB_13, Kumar_PRB_20}. However, a large discrepancy in the literature can be found regarding the effective magnetic moment and hence the spin states of Co$^{3+}$ in bulk Sr$_2$CoNbO$_6$. More recently, we found the effective magnetic moment of 4.6~$\mu_{\rm B}$, indicating the presence of Co$^{3+}$ predominantly in high spin (HS) state (d$^6$, t$_{2g}^4$e$_g^2$, S=2) \cite{Kumar_PRB_20}, as also reported in reference~\cite{Yoshii_JAC_2000}. Whereas significantly lower values of 1.91~$\mu_{\rm B}$ and 2.06~$\mu_{\rm B}$ were observed in references~\cite{Kobayashi_JPSJ_12} and \cite{Azcondo_Dalton_15}, respectively, and hence presence of LS (t$_{2g}^6$e$_g^0$, S=0)--HS or LS--IS (t$_{2g}^5$e$_g^1$, S=1) states indicate the strong temperature dependent spin-state transition in the underlined compound. The density functional theory (DFT) calculations within the generalized gradient approximation predict the semiconducting nature of the material with 0.2~eV band gap \cite{Ramirez_IJMPB_13}, which is supported by the positive temperature coefficient of conductivity carried out on bulk Sr$_2$CoNbO$_6$ with an activation energy of 0.25--0.35~eV, where conductivity sensitively increase with the oxygen partial pressure at 973~K \cite{Xu_JCSJ_16, Kumar_PRB_20}. Wang {\it et al.} showed the colossal dielectric properties in the compound, where the dielectric response was found to be closely related to the conduction mechanism governed by the hopping of localized charge carriers \cite{Wang_AIP_13}. Electronic band structure can give a valuable information about the spin state of Co$^{3+}$ and hence the magnetic properties of Sr$_2$CoNbO$_6$ as LS Co$^{3+}$ is insulating while IS/HS states are electrically conducting due to the presence of unpaired electrons. However, despite of the several interesting physical properties, the effect of biaxial strain on the physical properties of Sr$_2$CoNbO$_6$ thin films has not been explored.

In this paper we explore the substrate induced strain governed structural, transport, optical and electronic properties of Sr$_2$CoNbO$_6$ epitaxial thin films. The $\theta$-2$\theta$, $\phi$-scan, rocking curve and reciprocal space mapping (RSM) have been performed in order to extract both in-plane and out-of-plane information. The surface topography of the films analyzed using the atomic force microscopy (AFM) indicate an enhancement in the root mean square (RMS) roughness of the films with the misfit induced strain. We observe suppression (enhancement) in the electronic band gap with increase in the compressive (tensile) strain. Two distinct absorption bands have been observed in the visible-near infrared (Vis-NIR) absorption spectroscopy with the highest band gap in case of tensile strain. Further, we use x-ray photoemission spectroscopy (XPS) to measure Sr 3$d$, Nb 3$d$, Co 2$p$ and O 1$s$ core levels to understand the electronic properties and valence states, which play a crucial role in controlling the transport mechanism in these samples. An enhancement in the structural ordering in thin films as compared to the bulk is evident from the presence of several active modes in the Raman spectroscopy measurements.

\section{\noindent ~Experimental}

Polycrystalline target of Sr$_2$CoNbO$_6$ was synthesized by usual solid-state route. We use stoichiometric amount of strontium carbonate (SrCO$_{3}$, 99.995$\%$ Sigma-Aldrich), niobium (V) oxide (Nb$_{2}$O$_{5}$, 99.99$\%$ Sigma-Aldrich) and cobalt (II,III) oxide (Co$_{3}$O$_{4}$, 99.9985$\%$ Alfa Aesar) and mixed in a mortar-pestle for 5 hours for homogeneous mixing. Then we pressed the resultant powder into a pellet with the hydraulic press at 2500 psi and calcined at 900$^0$C for 2 hours and finally sintered at 1300$^0$C for 48 hours with several intermediate grindings \cite{Kumar_PRB_20}. Thin films from the circular bulk target of 20~mm diameter were grown by the  pulsed laser deposition (PLD) technique using 248~nm KrF excimer laser equipped with ultra high vacuum (UHV) growth chamber and a load-lock chamber \cite{Shukla_arXiv}. We have optimized growth parameters like deposition temperature, oxygen partial pressure, target to substrate distance, laser fluence and post deposition annealing temperature, oxygen pressure and time to get the best quality films. Finally the thin films of Sr$_2$CoNbO$_6$ were grown on LaAlO$_3$(100)  (LAO), (LaAlO$_3$)$_{0.3}$(Sr$_2$AlTaO$_6$)$_{0.7}$(100) (LSAT), SrTiO$_3$(100) (STO) and KTaO$_3$(110) (KTO) substrates ranging from compressive to tensile strain as shown in Fig.~\ref{Figure_0_misfit} (a), with the optimized parameters. A laser fluence of 1.5--2 Jule cm$^{-2}$ with 5~Hz pulse repetition rate and 5~cm target to substrate distance was used. All the films were grown at 800\degree C substrate temperature and 10$^{-3}$ mbar oxygen (99.99\%) partial pressure, starting from 9$\times$10$^{-10}$ mbar base pressure. We then annealed the films at 50~mbar oxygen partial pressure at the deposition temperature for 15~mins in order to maintain the oxygen stoichiometry. 
 
Phase purity of the bulk target sample was confirmed by the room temperature X-ray diffraction (XRD) measurements using PANalytical Aeris diffractometer in Bragg- Brentano geometry using CuK$\alpha$ ($\lambda$=1.5406 \AA) radiation with the accelerating voltage of 40 keV. FullProf suite with the linear background between two data points and pseudo voigt peak shape was used for the rietveld refinement of XRD data in order to extract the bulk lattice parameters of Sr$_2$CoNbO$_6$ \cite{Kumar_PRB_20}. Further, to confirm the epitaxy and quality of the films, $\theta$-2$\theta$, $\phi$-scan, rocking curve and RSM data were collected using PANalytical X'pert Pro MRD HR--XRD equipment. A stylus profiler was used to estimate the thickness of the films. For the thickness measurements, films with the sharp step were grown on Si(100) substrate by putting other piece of substrate on top of that to partially shadow the substrate. We used atomic force microscope (AFM) in the tapping mode to study the surface topography of the films. Temperature dependent resistivity measurements were carried out using  physical property measurement system (PPMS) of Quantum design, USA at 0.1 $\mu$A activation current. UV-Vis-NIR spectroscopy was carried out in the transmission mode using Perkin Elmer Lambda 1050 spectrometer. We measured core-level x-ray photoemission spectra using a monochromotic Al-K$\alpha$ (h$\nu=$ 1486.6 eV) source with a charge neutralizer. These spectra were fitted after the carbon correction and subtraction of inelastic Taugaard background using CasaXPS software considering both Gaussian and Lorentzian contributions. The Raman spectroscopic measurements were carried out using a Renishew inVia confocal microscope with 514.5~nm excitation wavelength and fitted with the Lorentizian function after the subtraction of the spline background between the data points. 

\section{\noindent ~Results and discussion}

 \begin{figure}
\includegraphics[width=3.0in]{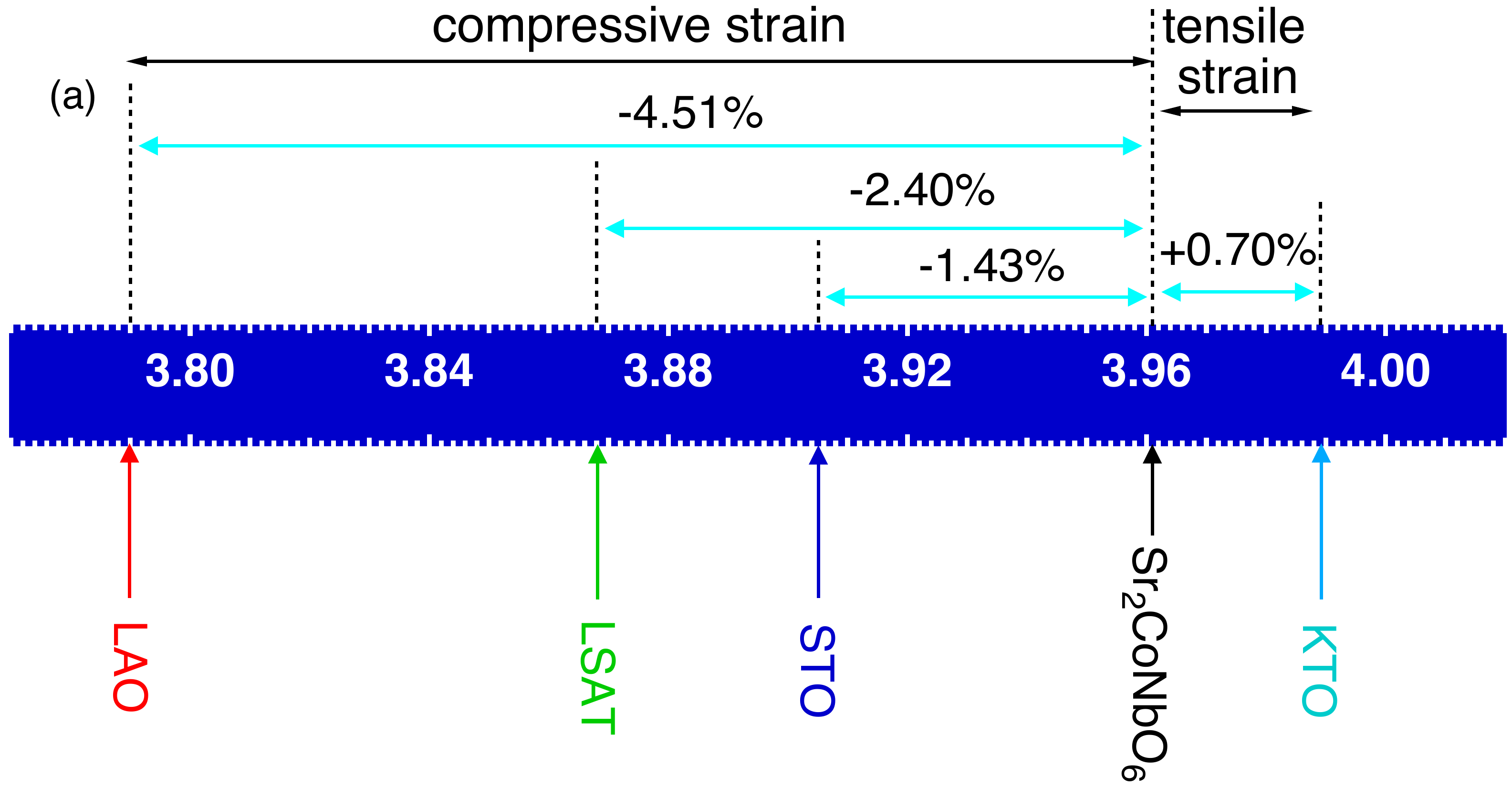}
\includegraphics[width=3.0in]{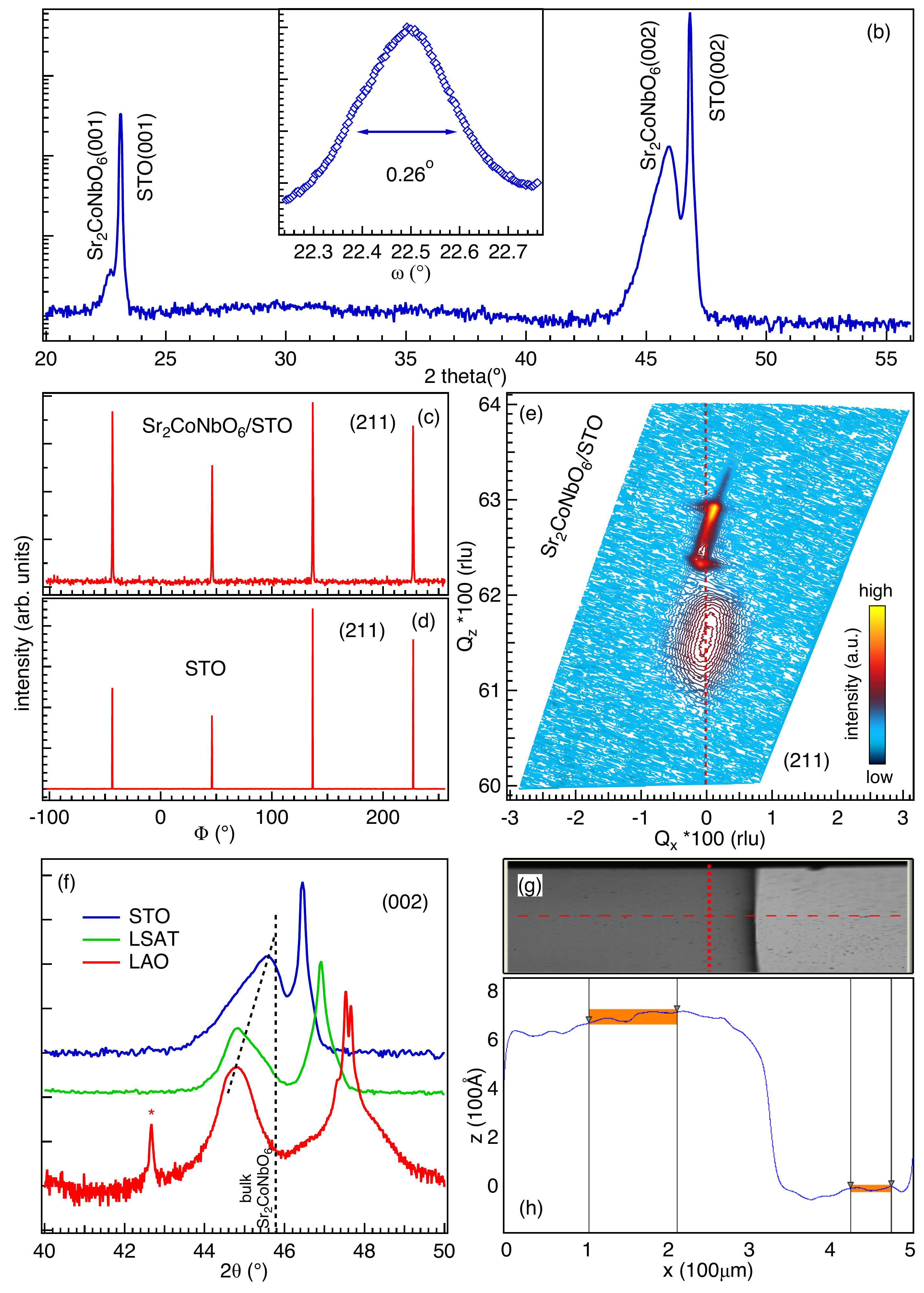} 
\caption {(a) Pseudocubic lattice parameters of the bulk target material Sr$_2$CoNbO$_6$ and different substrates used in this study with their in-plane lattice misfit, where misfit=(a$_{sub}$-a$_{bulk}$)/a$_{sub}$*100($\%$), a$_{sub}$ and a$_{bulk}$ are in-plane pseudocubic lattice parameters of substrates and bulk, (b) $\theta$-2$\theta$ XRD pattern of Sr$_2$CoNbO$_6$/STO film and inset shows the rocking curve about (002) reflection, (c) and (d) show the $\phi$-scan of the film and bare substrate, respectively, about (211) asymmetric reflection, (e) reciprocal space mapping (RSM) of Sr$_2$CoNbO$_6$/STO film about (211) reflection, (f) (002) reflection of the films on STO, LSAT and LAO substrates. $\star$ in case of LAO indicates the peak originating from the substrate, (g) and (h) show the step on the substrate surface and thickness measurement across the step from the stylus profilometer.}
\label{Figure_0_misfit}
\label{Figure_1_XRD}
\end{figure}

The XRD pattern of Sr$_2$CoNbO$_6$ thin film grown on STO substrate recorded in $\theta$--2$\theta$ mode is shown in Fig.~\ref{Figure_1_XRD}(b). The presence of only ({\it 00l}) reflections in the XRD pattern confirm the c-axis oriented growth of the film, where larger pseudo cubic lattice parameter of bulk Sr$_2$CoNbO$_6$ (3.961~\AA) as compared to STO substrate (3.905~\AA) gives rise to the reflection corresponding to the film at lower 2$\theta$ value. In the inset of Fig.~\ref{Figure_1_XRD}(b), we show the rocking curve about (002) reflection, which shows a full width at half maxima (FWHM) value of 0.26$^\circ$. Further, the $\phi$-scans were carried out around (211) asymmetric reflection to confirm the cube-on-cube epitaxy of the film with the substrate. Figures~\ref{Figure_1_XRD}(c) and (d) show the $\phi$-scan around (211) reflection for Sr$_2$CoNbO$_6$/STO film and bare STO substrate, respectively. Importantly, the observed four fold symmetry in case of both film as well as substrate indicates the epitaxy of the film with the substrate. Moreover, we perform the reciprocal space mapping (RSM) of Sr$_2$CoNbO$_6$/STO film about (211) reflection [as shown in Fig.~\ref{Figure_1_XRD}(e)], where Q$_x=$ 1/$\lambda$(cos $\omega$-cos(2$\theta$-$\omega$)) and Q$_z=$ 1/$\lambda$(sin $\omega$+sin(2$\theta$-$\omega$)) represent the in-plane and out-of-plane components of the scattering vector, respectively. Red dashed thin line in Fig.~\ref{Figure_1_XRD}(e) is to guide the eyes for the coherence between lattice parameters of the film and substrate. Here, two spots observed for the substrate may be due to the presence of twinning in the crystal. Figure~\ref {Figure_1_XRD}(f) shows the (002) reflection of Sr$_2$CoNbO$_6$ films grown on STO, LSAT and LAO substrates. Shift in the substrate peaks towards the higher 2$\theta$ value from STO to LAO can be clearly observed due to reduction in their lattice parameters [see Fig.~\ref{Figure_0_misfit}(a)], while that for thin films shows a shift towards the lower 2$\theta$ value as guided by black dashed line. An increase in the compressive in-plane lattice strain from STO to LAO results in the elongation of the out-of-plane lattice parameter of the films, and consequently a shift towards the lower 2$\theta$ value. A more prominent shift can be observed from STO to LSAT rather than LSAT to LAO, which may be due to increase in the degree of relaxation in the films in case of substrates with larger lattice misfit. From the $\theta$-2$\theta$ XRD patterns, we have extracted the out-of-plane lattice strain using $\varepsilon_{zz}$=(c$_{\rm film}$-c$_{\rm bulk}$)/c$_{\rm bulk}$*100($\%$), where c$_{\rm bulk}$ and c$_{\rm film}$ are out-of-plane pseudocubic lattice parameters of bulk and thin films of Sr$_2$CoNbO$_6$, which was found to be +0.35, +1.97 and +2.14$\%$, for STO, LSAT and LAO, respectively. In case of LAO substrate, a reflection peak at around 42.3 [marked by star in Fig.~\ref{Figure_1_XRD}(f)] originates from the substrate due to imperfection in the crystal. Fig.~\ref{Figure_1_XRD}(h) shows the stylus profile across the step on the film [as shown in Fig.~\ref{Figure_1_XRD}(g)] to estimate the thickness of the film, where the two horizontal strips indicate the average height taken over the scan before and after the step. We have estimated the film thickness of 70$\pm$5~nm by performing the measurements at two different places across the step for all the samples.

 \begin{figure}[h]
\includegraphics[width=3.35in]{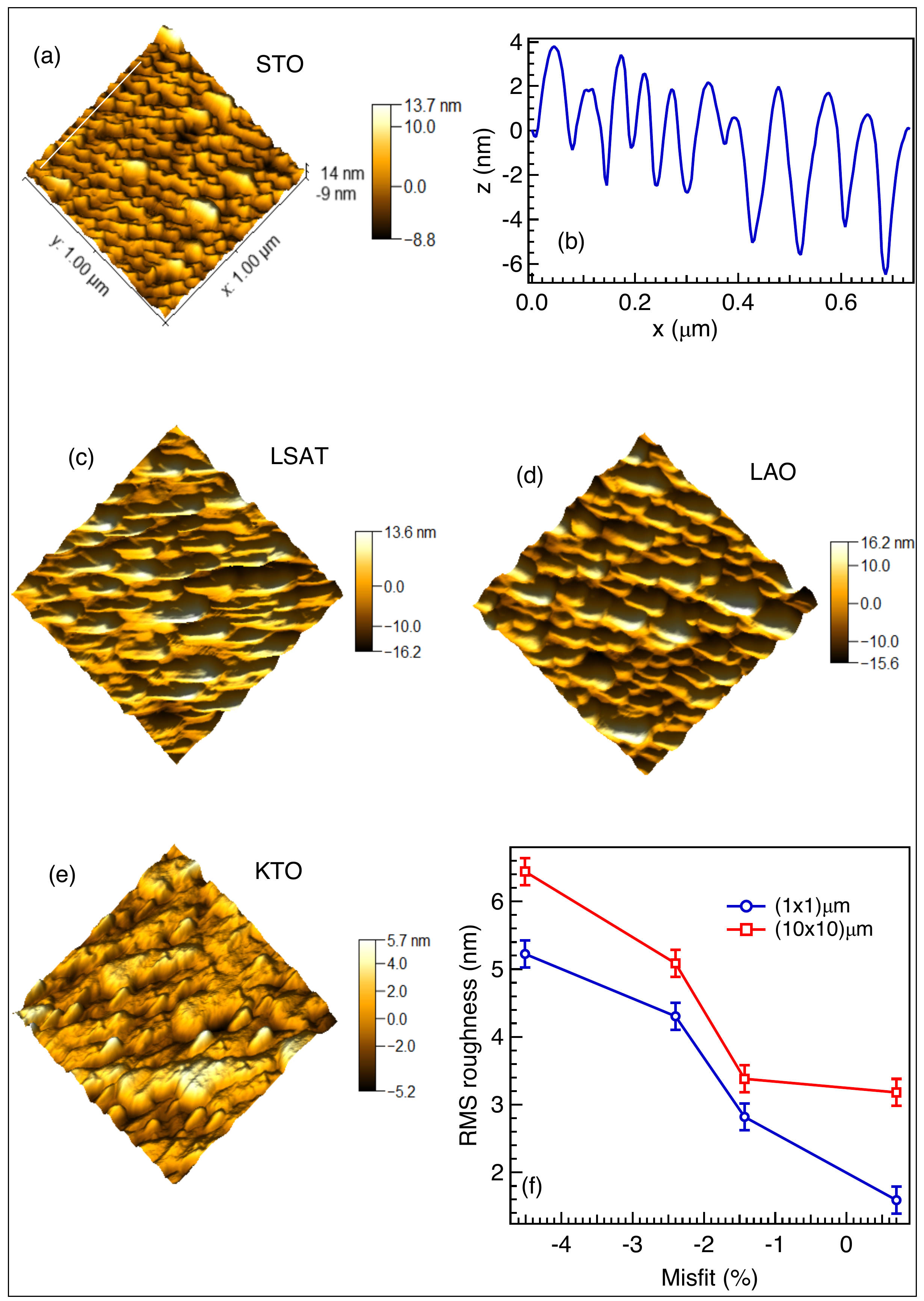}
\caption {(a) AFM image of (1x1) $\mu$m area of Sr$_2$CoNbO$_6$/STO film, (b) 1D scan across the periodic features of the film (white line in (a)), (c)--(e)  (1x1) $\mu$m AFM image of Sr$_2$CoNbO$_6$ film on LSAT, LAO and KTO substrates, respectively. (f) Dependence of RMS roughness of the films on the lattice misfit for (1x1) and (10x10) $\mu$m scan areas.}
\label{Fig2_AFM}
\end{figure}

Further, we have studied the surface topography of the thin film samples using the AFM imaging in tapping mode. Figures~\ref{Fig2_AFM}(a, c, d, e) show the AFM images of the films on STO, LSAT, LAO and KTO substrates, respectively, in (1x1) $\mu$m area. In Fig.~\ref{Fig2_AFM}(b), we plot a line profile of Sr$_2$CoNbO$_6$ film deposited on STO substrate, as shown by white line in Fig.~\ref{Fig2_AFM}(a), which shows the presence of periodic surface features on top of the films. It has been observed that the separation of these periodic features increases with the compressive strain, while these features are less prominent in case of tensile strain. Moreover, the RMS roughness of the films with change in the lattice misfit was extracted for (1$\times$1)~$\mu$m as well as (10$\times$10)~$\mu$m (not shown here) scan areas, as shown in Fig.~\ref{Fig2_AFM}(f). It can be clearly infer from the figure that roughness of the films increases with the lattice strain. The possibility of partial relaxation in the films, i.e., a loss of epitaxy can be expected for higher degree of lattice mismatch in LAO, which is also evident as the shift in the film peak corresponding to the out of plane lattice parameter as a consequence of the in-plane strain found to be a bit less in XRD of Sr$_2$CoNbO$_6$/LAO [see Fig.~\ref{Figure_1_XRD}(f)] as compared to the expected from the larger misfit in LAO, see the \% numbers in Fig.~\ref{Figure_1_XRD}(a) and reference \cite{Nichols_APL_13}. This can change the growth modes and consequently the grain size, which may result in the high roughness of the films with increase in the strain. 

\begin{figure}  
\includegraphics[width=3.35in]{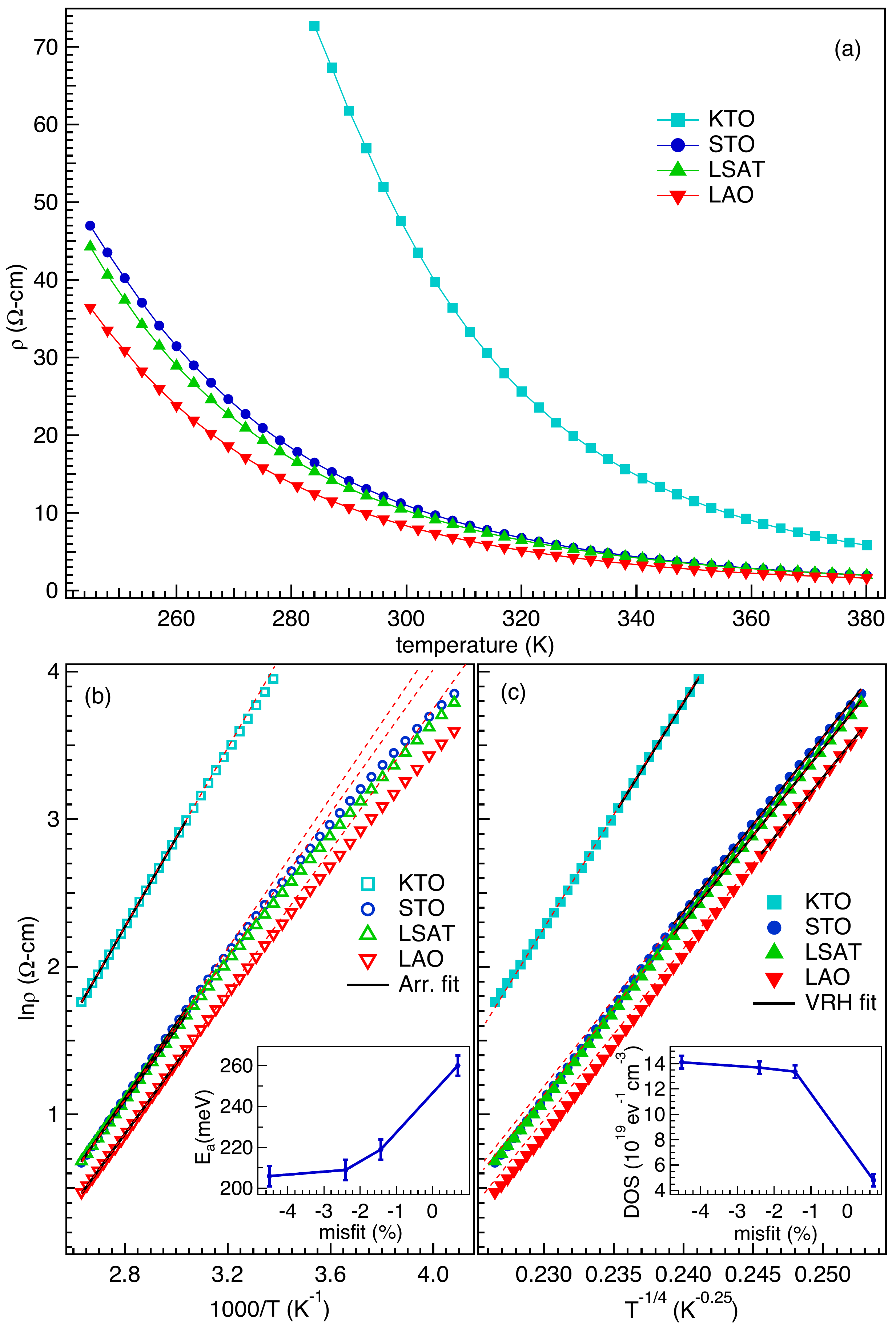}
\caption {(a) Temperature dependent resistivity of Sr$_2$CoNbO$_6$ films grown on different substrates. (b) and (c)  Arrhenius and VRH fitting in higher and lower temperature regimes, respectively. Insets show change in the activation energy and DOS with the lattice misfit, respectively. Red dashed lines indicate the extrapolation of the linear fitting to indicate the deviation from the Arrhenius and VRH models in the lower and higher temperature regimes respectively.} 
\label{Fig3_RT}
\end{figure}

In order to study the effect of the misfit induced biaxial strain on the electronic transport properties, we have performed temperature dependent resistivity ($\rho$--T) measurements on all the samples at 0.1~$\mu$A activation current, as shown in Fig.~\ref{Fig3_RT}(a). A negative temperature coefficient of resistivity indicate the semiconducting/insulating behavior of all the samples analogous to its bulk counterpart \cite{Xu_JCSJ_16, Kumar_PRB_20}. It can be clearly observed that resistivity of the samples decreases with increase in the compressive strain from STO to LAO while sharply increase in case of tensile strain (KTO). The conduction mechanism of the films in the high temperature regime is described by the Arrhenius model, which gives an estimation of the activation energy required by an electron to take part in the conduction mechanism. Resistivity can be described by the Arrhenius equation, as given below:
\begin{eqnarray}
 \rho(T)= \rho(0)exp(E_a/k_BT)
\end{eqnarray}
where E$_a$ and $\rho$(0) are the activation energy and pre-exponential factor, respectively. Activation energy extracted from the linear fitting of 1/T versus ln($\rho$) plot in high temperature region (330--380~K) is shown in the inset of Fig.~\ref{Fig3_RT}(b). A clear reduction in the activation energy in case of compressive strain, while a sharp enhancement in case of tensile strain is in well agreement with the decrease (increase) in the temperature dependent resistivity with the compressive (tensile) strain, as shown in Fig.~\ref{Fig3_RT}(a). However, the deviation from the Arrhenius model as shown by the extrapolated red dashed lines in Fig.~\ref{Fig3_RT}(b) indicate the presence of some other conduction mechanism in the lower temperature regime. We observe that conduction mechanism of the films in the lower temperature regime is governed by the Mott variable range hopping (VRH) model due to the localized charge carriers. Mott variable range hopping model describes the temperature dependence of resistivity as
\begin{eqnarray}
 \rho(T)= \rho(0)exp(T_0/T)^{1/4}
\end{eqnarray}
where T$_0$ is the characteristic temperature given as, T$_0$ = 18/k$_B$N(E$_F$)L$^3$, where N(E$_F$) and L are the localized density of states (DOS) near the Fermi level and localization length, respectively. A linear fitting of T$^{-1/4}$ versus ln($\rho$) in the lower temperature regime gives the characteristic temperature, T$_0$, which can be used to estimate the effective density of states near the Fermi level, by taking average Co--O ($\approx$2~\AA) bond length as the localization length, as shown in the inset of Fig.~\ref{Fig3_RT}(c). An increment in the effective DOS near the Fermi level with compressive strain, while reduction in case of tensile strain is consistent with the  change in activation energy and temperature dependent resistivity with the strain. A strong enhancement in the metal--ligand overlapping in case of compressive strain can be the possible reason for this enhancement in the electronic conductivity from STO to LAO while the suppression of the $p$--$d$ hybridization for tensile strain results in the sharp reduction in the electronic conductivity \cite{Huang_APL_19}. Further, it is important to re-emphasize that the effect of misfit induced strain and oxygen non-stoichiometry are strongly coupled with each other, as explicitly shown in the references \cite{Petrie_AFM_16, Herklotz_JPCM_17, Aschauer_PRB_13}. The tensile strain is widely known to favor the oxygen vacancies in the epitaxial films due to expansion in the unit cell volume \cite{Petrie_AFM_16, Herklotz_JPCM_17, Aschauer_PRB_13}. In the present case, in-plane tensile strain results in the expansion of the unit cell and hence favors the conversion of Co$^{3+}$ (as in bulk target material \cite{Xu_JCSJ_16, Yoshii_JAC_2000, Kumar_PRB_20}) to Co$^{2+}$ because of the large ionic radii of the later \cite{Shannon_AC_76}. At the same time, an enhancement in the concentration of insulating Co$^{2+}$ ions due to oxygen deficiency can also be the possible reason for this noteworthy reduction in the electronic conductivity in case of Sr$_2$CoNbO$_6$/KTO film sample. On the other hand, an increase in the compressive strain favors the conversion of Co$^{3+}$ to Co$^{4+}$ as well as strengthen the Co/Nb $d$ and O $p$ orbital overlapping, which give rise to enhancement in the conductivity in the films. However, it is difficult to quantitatively disentangle these effects on the conduction mechanism of these films.

\begin{figure}[h]
\includegraphics[width=3.35in]{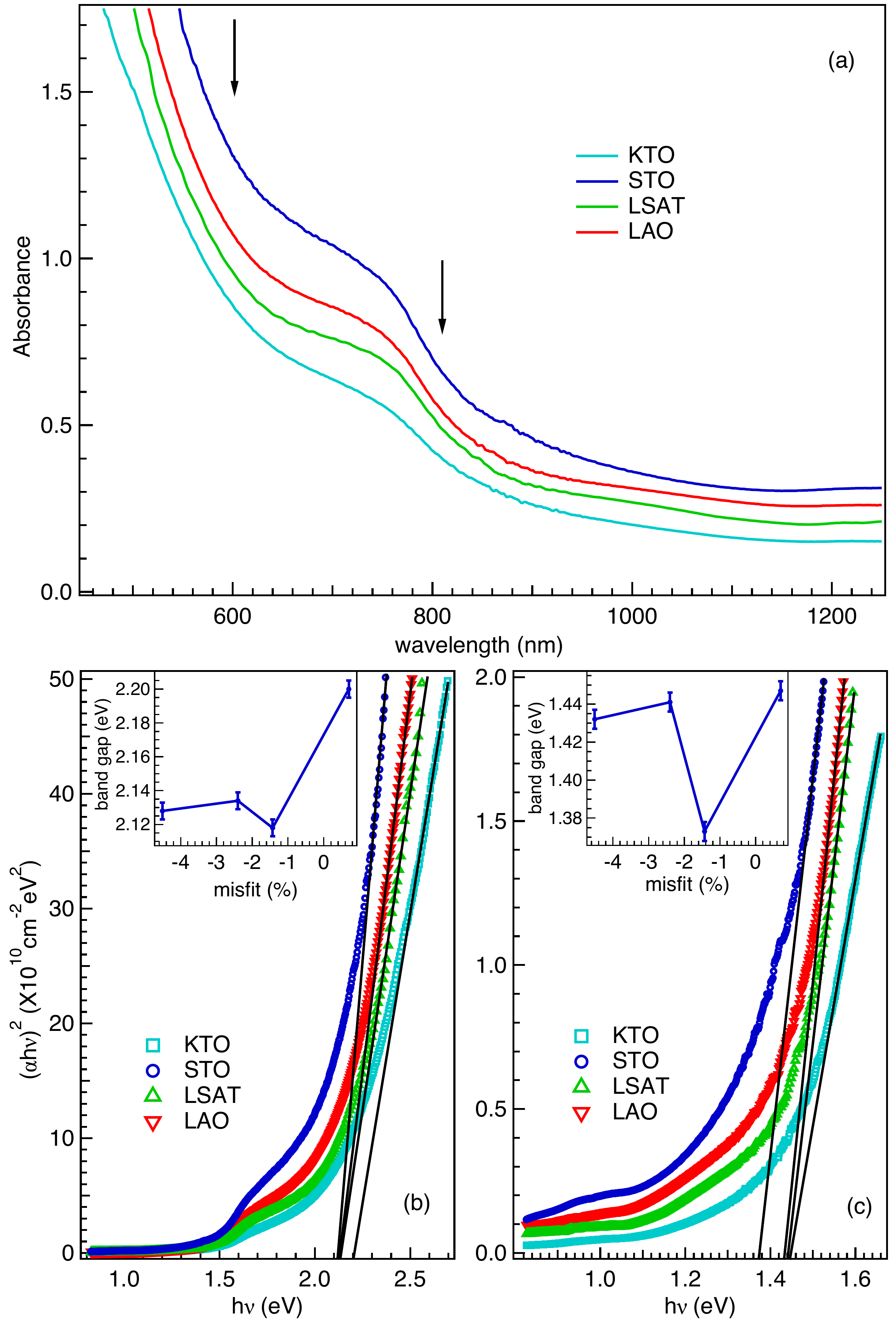}
\caption {(a) Room temperature absorption spectra in visible and NIR region for Sr$_2$CoNbO$_6$ films grown on various substrates. (b) and (c) the Tauc plots in two different energy regimes and insets show the variation of the band gaps with the lattice misfit, estimated from the respective Tauc plot.} 
\label{Figure_8_UV}
\end{figure}

Interestingly, strain induced electronic band gap engineering further facilitate to probe the effect of biaxial strain on the optical band gap of these samples. The Vis-NIR absorption spectroscopy measurements clearly show two distinct absorption bands around 600 and 800 cm$^{-1}$ [indicated by arrows in Fig.~\ref{Figure_8_UV}(a)]. These two bands in the optical and near IR regions can be due to the charge transfer between O 2$p$ and Co 3$d$/Nb 4$d$ orbitals because of the covalent character of the bonding and intraband transitions within the Co 3$d$/Nb 4$d$ orbitals \cite{Huang_APL_19}. In the present case, we can extract a direct band gap using the following equation \cite{Tauc_1974}
\begin{eqnarray}
(\alpha h \nu)^2= A(h\nu -E_g)
\end{eqnarray}
where $\alpha$, h$\nu$, A and E$_g$ are the absorption coefficient, photon energy, proportionality constant and direct band gap, respectively. The estimated values of band gap from the linear extrapolation of the absorption region in ($\alpha$h$\nu$)$^2$ versus h$\nu$ plot (Tauc plot) are shown in the inset of Figs.~\ref{Figure_8_UV}(b, c). We note here that both the optical band gaps show the highest values in case of tensile strain (Sr$_2$CoNbO$_6$/KTO sample), analogous to the electronic band gap [see inset of Fig.~\ref{Fig3_RT}(b)]. On the other hand, in case of the compressive strain the higher band gap corresponding to the absorption band at $\sim$600~nm does not show the significant variation, whereas that corresponding to $\sim$800~nm shows a significant reduction in case of Sr$_2$CoNbO$_6$/STO sample, unlike the electronic band gap, which shows a monotonic decrease with the compressive strain. This discrepancy in the behavior of optical and electronic band gap in case of compressive strain needs the further investigation by performing the band structure calculations \cite{Uba_PRB_12}.

\begin{figure}[h]  
\includegraphics[width=3.4in]{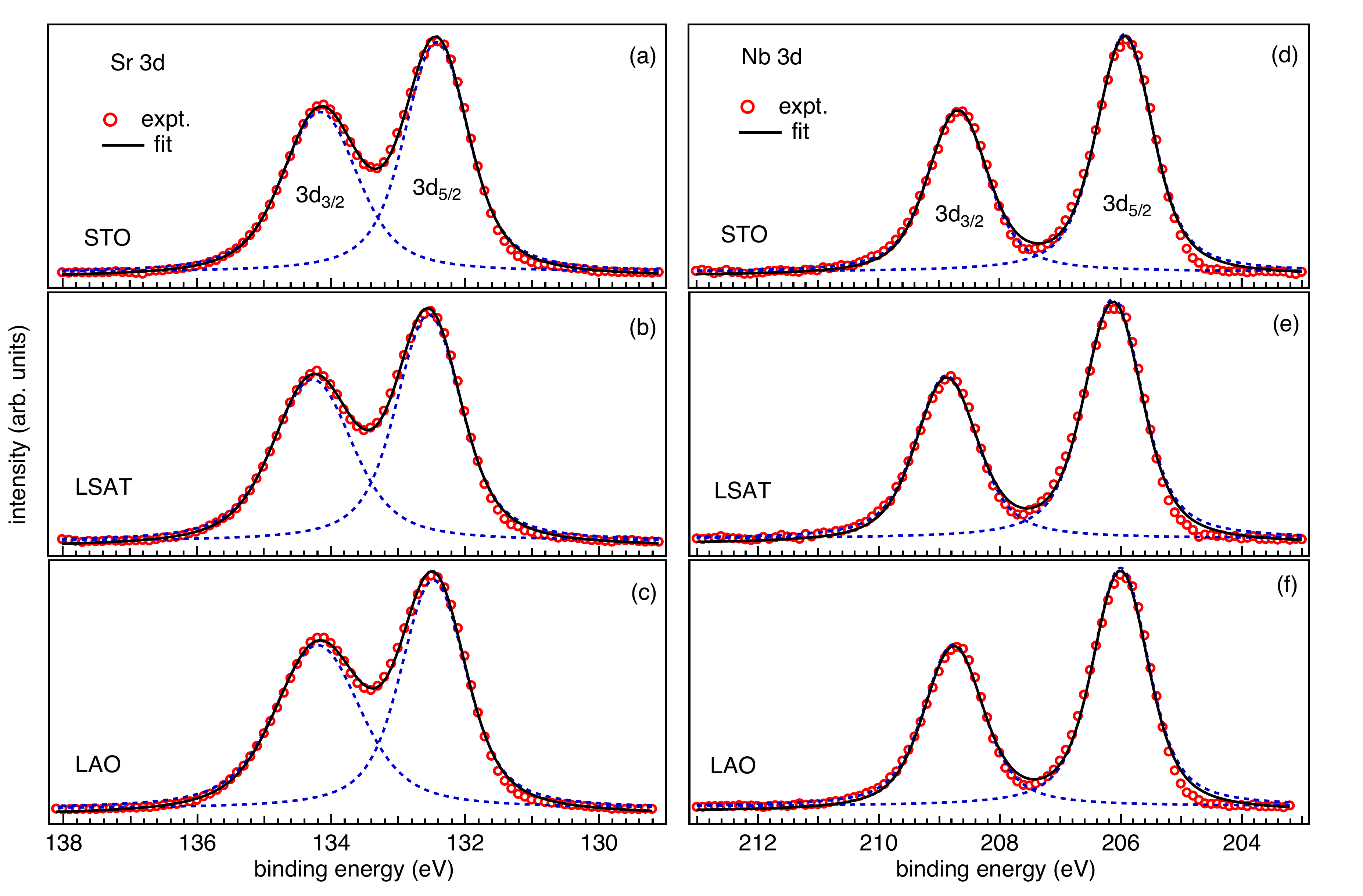}
\caption {The core-level spectra of Sr 3$d$ (a--c) and Nb 3$d$ (d--f) for Sr$_2$CoNbO$_6$ films grown on STO, LSAT and LAO substrates, respectively. The experimental data points and fitted curve are shown by red circles and black solid lines, respectively along with the deconvoluted spin-orbit components by blue dotted lines.} 
\label{Figure_3_XPS}
\end{figure}

In order to understand this strain induced change in the electronic  structure, we have measured core-levels of Sr 3$d$, Nb 3$d$, Co 2$p$ and O 1$s$ for Sr$_2$CoNbO$_6$ films grown on STO, LSAT and LAO substrates using x-ray photoelectron spectroscopy (XPS). Figs.~\ref{Figure_3_XPS}(a--c) show the Sr 3$d$ core-levels where  two peaks i.e 3$d_{5/2}$ and  3$d_{3/2}$ can be clearly seen around $\sim$ 132.5 and 134.2~eV due to spin-orbit splitting. These peak positions are in close agreement with the reported values for SrO \cite{Vasquez_JESRP_91, Sosulnikov_JESRP_92}, which confirms the presence of divalent Sr cations in the samples. Interestingly, we observe Nb 3$d_{5/2}$ and 3$d_{3/2}$ components around $\sim$ 206.0 and 208.8~eV, respectively [Figs.~\ref{Figure_3_XPS}(d--f)], which are close to the reported values for Nb$^{4+}$ \cite{Bahl_JPCS_75,Wong_PRB_14}, contradictory to the bulk Sr$_2$CoNbO$_6$, where pentavalent Nb is expected from the magnetization measurements as Co solely govern the magnetic properties in the bulk samples \cite{Kobayashi_JPSJ_12, Yoshii_JAC_2000, Kumar_PRB_20}. The presence of Nb in 4+ valence state in these thin film samples and hence an unpaired d-electron (4d$^1$) can interestingly govern the magnetic and transport properties in Sr$_2$CoNbO$_6$ thin films. Here, the observation of tetravalent Nb indicate the presence of either oxygen deficiency or lesser formal charge on the oxygen (presence of O$^{-1}$) due to the strong covalent character of the Co/Nb--O bond. We found that there is no significant shift in the peak positions of Sr 3$d$ and Nb 3$d$ core-levels, which discard any significant change in the valence state with substrate induced strain.

\begin{table*}
\centering
		\label{tab:Oxygen}
		\caption{The fitting parameters of O 1$s$ core-level spectra and the extracted surface composition of the different ions of Sr$_2$CoNbO$_6$ thin films deposited on various substrates.}

\begin{tabular}{p{2.5cm}p{1.8cm}p{1.8cm}p{1.8cm}p{1.8cm}p{2.4cm}p{3cm}}
		\hline
\hline
	 Substrate	& Oxygen  &FWHM & Peak area & $\frac {\rm{Area} \rm{(O^{1-})}}{\rm{Area}\rm{(O^{2-})}}$ &Average Oxygen &Sr:Co:Nb:O  \\
&components&(eV)& ($\%$)&&valency&($\pm$0.05) \\
\hline
 STO &O$_{\rm I}$& 1.2&67.4 &0.36 & -1.73&2.12:0.94:0.96:5.98\\
 &O$_{\rm II}$& 1.9&24.5 & & &\\
 &O$_{\rm III}$&2.1&8.1 & & &\\
LSAT &O$_{\rm I}$ &1.2& 52.9 &0.44 & -1.69 &2.08:0.94:0.96:6.02\\
 &O$_{\rm II}$ &1.9& 22.5 & &  &\\
 &O$_{\rm III}$ & 2.2& 24.6 && & \\
LAO &O$_{\rm I}$ &1.1& 65.3 &0.46 & -1.68&2.33:0.93:0.94:5.80 \\
 &O$_{\rm II}$ &2.1&29.7& & & \\
 &O$_{\rm III}$ &2.0&5.0 & & & \\
\hline
\hline
\end{tabular}
\end{table*}

\begin{figure}[h]  
\includegraphics[width=3.4in]{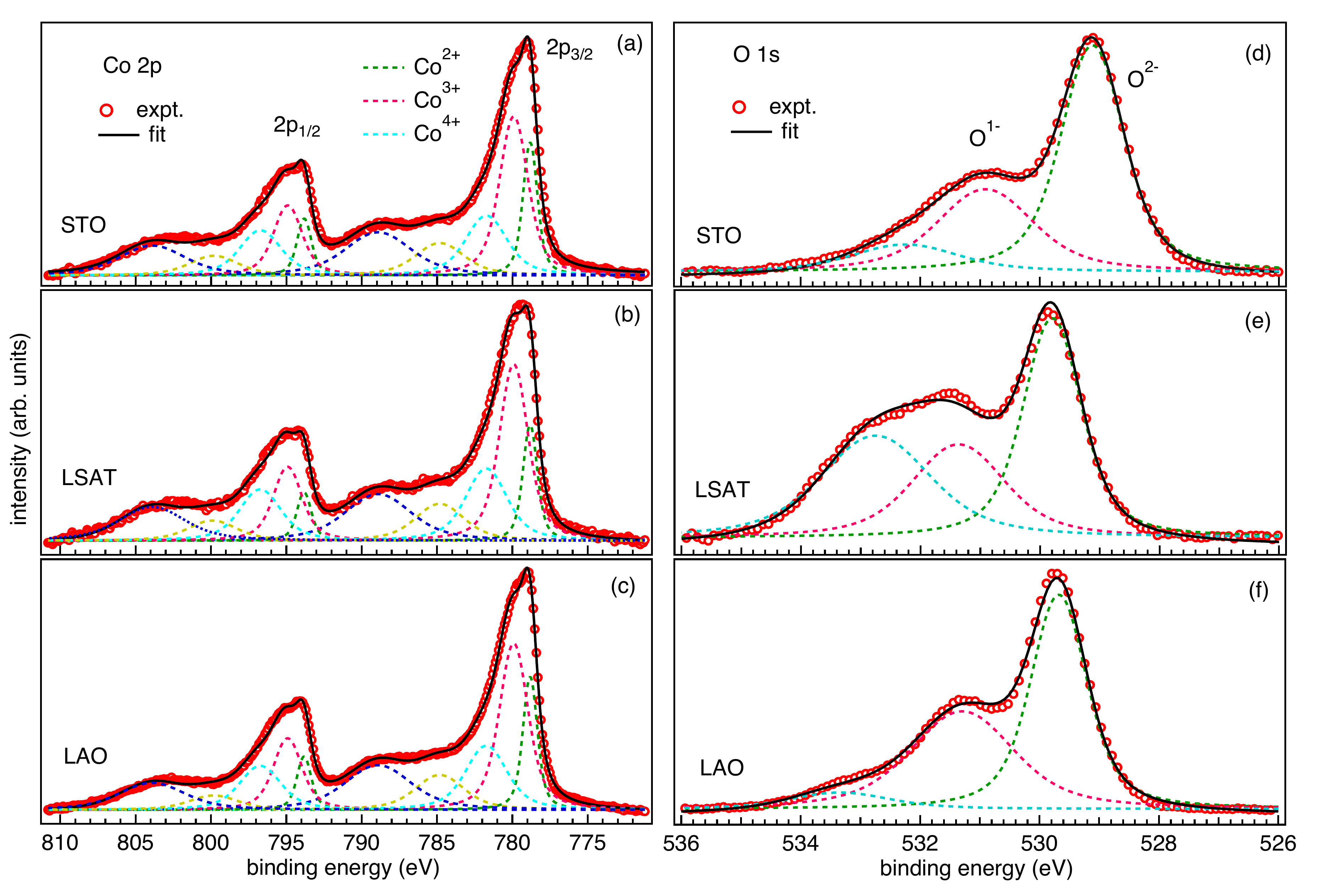}
\caption {The Co 2$p$ (a--c) and O 1$s$ (d-f) core-level spectra of Sr$_2$CoNbO$_6$ films deposited on different substrates.} 
\label{Figure_4_XPS}
\end{figure}

Figures~\ref{Figure_4_XPS}(a--c) show the Co 2$p$ core level spectra where two spin-orbit splitting components 2$p_{1/2}$ and 2$p_{3/2}$, are separated by $\sim$15~eV, which is consistent with the reported value \cite{Chung_SS_76}. These main peaks (2$p_{1/2}$ and 2$p_{3/2}$) are deconvoluted into three components, which can be assigned for the three oxidation states of Co i.e Co$^{2+}$, Co$^{3+}$ and  Co$^{4+}$. The peak position for each component are found to be at 778.5~eV, 780~eV, and 781.7~eV binding energies, which are consistent with reported values in references \cite{Chung_SS_76, Biesinger_ASS_11, Dupin_TSF_76}. The satellite feature around 784.8~eV i.e $\sim$6~eV above the main peak of Co$^{2+}$ and 788.9~eV i.e $\sim$9~eV above the main peak of Co$^{3+}$ are close to the reported values for CoO and Co$_3$O$_4$, respectively \cite{Chung_SS_76}. Unlike the bulk Sr$_2$CoNbO$_6$ (where Co is expected to be in 3+ state only \cite{Xu_JCSJ_16, Yoshii_JAC_2000, Kumar_PRB_20}), different valence states of Co in case of thin films can interestingly govern the related physical properties. Notably the compressive strain causes a reduction in the in-plane and consequently elongation in the out-of-plane lattice parameters \cite{Vailionis_PRB_11}. As a result of which Co$^{3+}$  prefers to stabilize in 4+ and 2+ states in the in-plane and out-of-plane Co--O bonds, owing to their significantly smaller and larger ionic radii in the two cases, respectively, as compared to Co$^{3+}$ \cite{Shannon_AC_76}. This could be the possible reason for the evolution of the 2+ and 4+ valence states of Co along with the dominating 3+ state observed in the XPS core-levels. A similar effect was observed in references \cite{Aschauer_PRB_13, Petrie_AFM_16} where strain was found to govern the different oxidation states of transition metal cations and hence oxygen vacancies in the compounds due to change in their molar volume.

Moreover, we have recorded the O 1$s$ core level spectra of Sr$_2$CoNbO$_6$ films deposited on STO, LSAT and LAO substrates, as shown in Figs.~\ref{Figure_4_XPS}(d--f), respectively. We have fitted each spectrum with three components and their fitting parameters are listed in the Table~I. The most intense peak around 529.7~eV (O$_{\rm I}$) is attributed to the lattice oxygen with 2- formal charge while the one with lowest intensity on the higher binding energy side around 533.0~eV (O$_{\rm III}$) corresponds to the surface contamination i.e hydroxylation and/or carbonation of the surface species. A huge discrepancy in the literature can be found regarding the origin of this central component of the O 1$s$ spectrum at around 531.4~eV (O$_{\rm II}$) \cite{Pawlak1_JPCB_99, Pawlak2_JPCB_99, Pawlak_JPCB_02}. Pawlak {\it et al.} have proposed a model describing the non-singularity of O 1$s$ peak in case of Y$_3$Al$_5$O$_{12}$ \cite{Pawlak1_JPCB_99} and YAlO$_3$ \cite{Pawlak2_JPCB_99}, where oxygen is surrounded by two different cations and then successfully applied to the perovskites of type (AA$^\prime$)(BB$^\prime$)O$_3$ \cite{Pawlak_JPCB_02}, where four cations of different covalent character are attached to the oxygen anions. This difference in the covalent character of the cations lead to a change in the valency of the oxygen anions in the complex compounds. Dupin {\it et al.} have reported the peaks around 531.1--532~eV due to the presence of O$^{1-}$ species due to low electronic concentration in the particular region because of the high covalent character of the transition metal--oxygen bond and describe these as low coordinated oxygen sites at the surface region as compared to O$^{2-}$ \cite {Dupin_PCCP_2000}. Interestingly, Wu {\it et al.} follow the same idea and successfully explained the average valency of oxygen (-1.55) in BaTiO$_3$ \cite{Wu_AIP_15}, which was found to be close to the theoretically reported value (-1.63) \cite{Cohen_Nature_92}.

We have also assigned this central component as O$^{1-}$ species and average valency of the oxygen anion estimated from the integrated intensity of O$^{2-}$ and O$^{1-}$ species, as given in Table~I, where we neglect the contribution from the O$_{\rm III}$ component \cite{Wu_AIP_15}. As speculated from Nb 3$d$ and Co 2$p$ spectra, the average valency of the oxygen is lower than 2, which indicate the presence of holes in the O 2$p$ orbital and estimation of which can be a useful tool in order to understand several physical properties in the underlined compounds \cite{Xu_PSSB_15}. The decrease in valency of oxygen anion indicate the enhancement in the covalent character of Co/Nb--O bonds due to the reduction in the Co/Nb--O bond length in case of compressive strain, which give rise to the enhancement in the delocalization of the electronic wave function. This enhancement in the concentration of delocalized electrons can be due to the increase in the electronic bandwidth with increase in the compressive strain as evident from the $\rho$--T measurements in Fig.~3. Further, the surface composition of the different ions in Sr$_2$CoNbO$_6$ films deposited on various substrates is estimated from the integrated intensities of their respective core levels and considering the corresponding photo-ionization cross section \cite{Yeh_Table_85, Dhaka_SS_09}, as summarized in the Table~I. We found that the films grown on STO and LSAT substrates are stoichiometric along with oxygen (within the sensitivity of XPS), whereas that on LAO are slightly oxygen deficient. This change in the oxygen stoichiometry could also play a crucial role in determining the conductivity of the samples. A cumulative effect of the change in metal-ligand hybridization due to strain and the presence of oxygen deficiency are expected to control the conduction mechanism with change in the substrate induced strain.

\begin{figure}[h]
\includegraphics[width=3.3in]{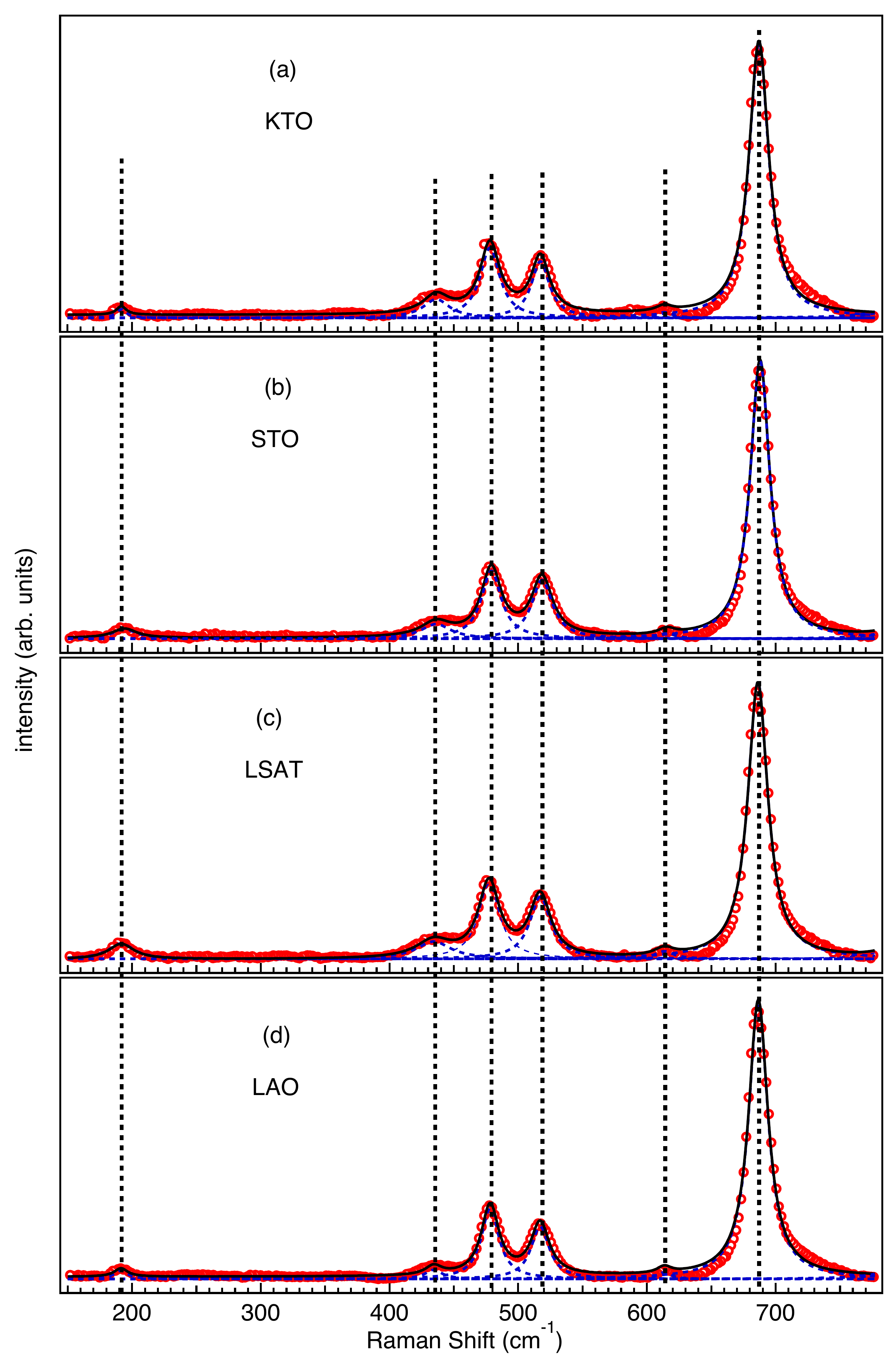}
\caption {Room temperature unpolarized Raman spectra of Sr$_2$CoNbO$_6$ thin films grown on various substrates.} 
\label{Figure_7_Raman}
\end{figure}

Finally, we study the vibrational energy levels using Raman spectroscopy, which is another important tool to understand the detailed structure including the octahedral distortion, rotation and/or tilting. The octahedral distortion lifts the degeneracy of the vibrational levels and degree of distortion decide the separation in the energy levels and hence splitting of the Raman modes, which is not possible using the conventional XRD method due to its low sensitivity for the lower Z elements like oxygen. Figures~\ref{Figure_7_Raman}(a--d) show the unpolarized room temperature Raman spectra, measured with 514.5~nm excitation wavelength. The presence of several Raman active modes for all the samples indicate a significant enhancement in the structural ordering in these epitaxial thin films as compared to the bulk, where only two weak Raman modes were detected in the bulk Sr$_2$CoNbO$_6$ sample \cite{Kumar_PRB_20}. A large discrepancy in the literature can be found regarding the crystal structure of bulk Sr$_2$CoNbO$_6$. It has been reported that the Sr$_2$CoNbO$_6$ and related compounds crystalize in cubic phase with the random occupancy of B-site cations i.e. perovskite unit cell with $Pm\bar{3}m$ space group \cite{Yoshii_JAC_2000, Azcondo_Dalton_15, Huang_CM_09}, while others have shown the rhombohedral structure with $R\bar{3}m$ space group \cite{Kobayashi_JPSJ_12} as well as octahedral tilted tetragonal structure with $I4/m$ space group having a$^{0}$a$^{0}$c$^{-}$ Glazer notation i.e antiphase tilting of NbO$_6$ and CoO$_6$ octahedra along [001] direction \cite{GlazerAC72, Wang_AIP_13, Bashir_SSS_11, Ramirez_IJMPB_13}. It is well known that  perovskites with ideal cubic structure ($Pm\bar{3}m$ with the random occupancy of B-site cations) is centrosymmetric and hence Raman inactive, while group theory predict four Raman active modes for cubic B-site ordered structure ($Fm\bar{3}m$ double perovskite structure, where unit cell doubles in all the three crystallographic directions) i.e. $\Gamma_g$ (Fm$\bar{3}$m) = $\nu_1$(A$_{1g}$) + $\nu_2$(E$_{g}$) + $\nu_5$(F$_{2g}$) + T(F$_{2g}$), where $\nu_1$ are the oxygen symmetric stretching modes along Co--O--Nb axis, $\nu_2$ are the two fold degenerate oxygen asymmetric stretching modes, where four oxygen anions in the same plane moves towards the octahedral center and remaining two perpendicular anions moves outward or vice-versa, $\nu_5$ are the triply degenerate oxygen bending modes and T indicate the modes resulting from the translational motion of Sr cations \cite{Iliev_PRB_07, Andrews_Dalton_15}. The presence of six Raman active modes for all the samples in the present case indicate the further lowering in the crystal symmetry from the ideal face-centered-cubic $Fm\bar{3}m$ structure due to octahedral rotation in case of thin films. Further, nine Raman active modes are predicted for the $I4/m$ crystal symmetry having the irreducible representation as $\Gamma_g$ (I4/m) = $\nu_1$(A$_{g}$) + $\nu_2$(A$_{g}$ + B$_g$) + $\nu_5$(B$_g$ + E$_{g}$) + T(B$_g$ + E$_{g}$) + L(A$_g$ + E$_g$), where, L denote the librational modes associated with the rotation of the (B/B$^\prime$)O$_6$ octahedra \cite{Andrews_Dalton_15, Guedes_JAP_07}. Highest intensity mode around $\sim$690~cm$^{-1}$ can be ascribed to the A$_{g}$ oxygen symmetric stretching mode due to its stiffer force constant, a low intensity mode around $\sim$610~cm$^{-1}$ is oxygen asymmetric stretching modes in which A$_{g}$ and B$_{g}$ modes can not be resolved due to its low intensity. Three Raman active modes of significant intensity between 400 to 550 cm$^{-1}$ are expected to be the oxygen bending modes, $\nu_5$, while only two such modes have been predicted for I4/m symmetry, suggesting further lowering in the crystal symmetry in epitaxial thin films due to substrate induced strain.

   A further a$^-$a$^-$b$^+$  tilting of octahedra can transform the system into monoclinic phase having $P2_1/n$ space group with 24 Raman active modes, represented as $\Gamma_g$ (P2$_1$/n) = $\nu_1$(A$_{g}$ + B$_g$) + 2$\nu_2$(A$_{g}$ + B$_g$) + 3$\nu_5$(A$_g$ + B$_{g}$) + 3T(A$_g$ + B$_{2g}$) + 3L(A$_g$ + E$_g$). A similar crystal symmetry was also observed in case of bulk A$_2$CoNbO$_6$ (A = Ba, Ca) compounds of the same family \cite{Bashir_SSS_11, Shaheen_SSS_10}. The presence of two formula units per unit cell give rise to both A$_g$ and B$_g$ splitting in all the modes. Thus, three Raman modes between 400--550 cm$^{-1}$ can be attributed to the oxygen bending modes and low intensity mode around 190 cm$^{-1}$ is the external T mode, which involve the movement of Sr atoms. Six T modes (3 A$_g$ and 3 B$_g$) are predicted from the group theory, but low intensity modes of background level, overlapping of the modes or presence of the modes below the spectrometer range (100 cm$^{-1}$) can be the possible reasons for the absence of the remaining T modes. The L modes are silent in the present case and A$_g$ and B$_g$ modes can not be resolved in any case within the instrumental resolution. Also, there is no significant shift in the peak position of the observed Raman modes with strain. Therefore, a polarization dependent Raman spectroscopy study is needed to get further insight into the structure and to extract the extent of Co/Nb ordering in the lattice.

\section{\noindent ~Conclusion}

We investigate the structural, transport, optical and electronic properties of the Sr$_2$CoNbO$_6$ thin films deposited on various substrates. The high resolution $\theta$--2$\theta$, $\omega$, $\phi$--scans and reciprocal space mapping indicate the growth of epitaxial and high-quality films. The temperature dependent resistivity measurements indicate the semiconducting/ insulating behavior of all the films. A systematic reduction in the electronic resistivity with the compressive strain while a sharp enhancement in case of tensile strain was observed, which further supported by the decrease (increase) in activation energy and increase (decrease) in the density of states (DOS) with the compressive (tensile) strain. The Vis-NIR spectroscopy measurements show two distinct absorption bands near 600 and 800 cm$^{-1}$ and corresponding band gaps show monotonic behavior with increase in the compressive strain while a sharp enhancement in case of the tensile strain. The photoemission study indicates the presence of tetravalent Nb and different oxidation states of Co for all the films under compressive strain. An increase in delocalized electrons which is evident from the enhancement in O$^{1-}$ species in O 1$s$ core-levels and the oxygen non-stoichiometry are the possible factors for the enhancement in the electronic conductivity in case of compressive strain. The presence of the sharp Raman active modes in all the films suggest the significant enhancement in the structural ordering in case of epitaxial thin films as compared to the bulk Sr$_2$CoNbO$_6$ sample. 

\section{\noindent ~Acknowledgments}

This work was financially supported by SERB-DST through Early Career Research (ECR) Award (project reference no. ECR/2015/000159) and planning unit of IIT Delhi through Seed Grant (reference no. BPHY2368). AK and RS gratefully acknowledge the UGC and DST-Inspire, India for fellowship. Authors acknowledge the central and nano research facilities of IIT Delhi for providing the research facilities: XRD, AFM, stylus profilometer, PPMS EVERCOOL-II, UV-Vis-NIR and Raman. We thank Ploybussara Gomasang for help in the XPS measurements, which were supported by Sakura Science Program (aPBL) at Shibaura Institute of Technology, Japan. We also thank Dr. Mahesh Chandra and Mr. Guru Dutt Gupt for useful discussion and help. A high temperature furnace from Nabertherm, supported by BRNS through DAE Young Scientist Research Award project sanction no. 34/20/12/2015/BRNS, was used for target preparation.



\end{document}